# gggenomes: effective and versatile visualizations for comparative genomics

Thomas Hackl[1,†,*], Markus Ankenbrand[2,†], Bart van Adrichem[1], David Wilkins[3], Kristina Haslinger[4]

1. Groningen Institute for Evolutionary Life Sciences, University of Groningen, Nijenborgh 7, 9747 AG Groningen, The Netherlands
2. Julius-Maximilians-Universität Würzburg, Fakultät für Biologie, Center for Computational and Theoretical Biology, Klara-Oppenheimer-Weg 32, 97074 Würzburg, Germany (markus.ankenbrand@uni-wuerzburg.de)
3. Discipline of General Practice, Adelaide Medical School, The University of Adelaide, North Terrace, Adelaide, South Australia 5005, Australia (david@wilkox.org)
4. Department of Chemical and Pharmaceutical Biology, Groningen Research Institute of Pharmacy, University of Groningen, Antonius Deusinglaan 1, 9713 AV Groningen, The Netherlands (k.haslinger@rug.nl)

†These authors contributed equally to this work
*Correspondence: t.hackl@rug.nl

The effective visualization of genomic data is crucial for exploring and interpreting complex relationships within and across genes and genomes. Despite advances in developing dedicated bioinformatics software, common visualization tools often fail to efficiently integrate the diverse datasets produced in comparative genomics, lack intuitive interfaces to construct complex plots and are missing functionalities to inspect the underlying data iteratively and at scale. Here, we introduce *gggenomes*, a versatile R package designed to overcome these challenges by extending the widely used *ggplot2* framework for comparative genomics. *gggenomes* is available from CRAN and GitHub, accompanied by detailed and user-friendly documentation (https://thackl.github.io/gggenomes).

*gggenomes* builds on the *ggplot2* framework and the *tidyverse* philosophy, two essential components for intuitive, flexible plotting in R. *ggplot2*, one of the most widely used visualization packages, employs a "grammar of graphics" that constructs plots through a modular layering approach, combining elements such as axes, scales, and labels. This modular structure, which enables building complex visualizations in a straightforward, highly customizable manner, also provides an ideal platform for visualizing and integrating diverse genomic data. *gggenomes* expands on *ggplot2*'s capabilities with dedicated visuals for chromosomes, genes, synteny blocks, and other genomic features, and by seamlessly supporting hierarchical genomic coordinates organized in genomes, chromosomes, contigs, and features such as genes annotated on these primary sequences. Through this flexibility, *gggenomes* supports both rapid diagnostic plots during data analysis and polished, publication-ready visuals for dissemination.

Grounded in the *tidyverse* principles (https://tidyverse.tidyverse.org/articles/manifesto.html), *gggenomes* further simplifies data integration by working directly with tidy data tables. Unlike many genomics tools that require specific, rigid file formats, *gggenomes* offers built-in importers for widely used formats like FASTA, GenBank, GFF, BED, VCF, BLAST, and PAF. Additionally, it supports generic tables and custom datasets, read in via other R packages or custom parsers, which allows researchers to visualize an extensive range of genomic data without complex transformations. This tidy approach not only streamlines integration across data types but also allows researchers to enrich their plots with other *ggplot2*-based packages and annotations, creating a cohesive visualization environment throughout the research process.

The under-the-hood design of *gggenomes* is inspired by several other R packages. Its visualization of genes is based on *gggenes* (https://wilkox.org/gggenes), extended with the ability to display intron-containing gene models. Its approach to integrating diverse datasets draws from *ggraph and ggtree*[1] and led to one of the package's key concepts compared to other *ggplot2*-based packages: a lightweight track system that allows users to seamlessly visualize multiple datasets simultaneously. This feature is essential for comparative genomics, where displaying complementing information in data layers – such as gene annotations, sequence motifs, gene expression, synteny relations and homology information – is necessary for being able to contrast signals, recognize patterns, and ultimately uncover underlying relationships.

Another key strength of *gggenomes* is its capacity for scripted interactivity and downstream data export, enhancing scalable data exploration and analysis. Users can import datasets ranging from single regions to entire annotated genomes yet selectively visualize only regions of interest based on metadata criteria specified at the time of plotting. These criteria can be flexibly chosen from any available data track, enabling complex operations such as zooming into "all gene clusters with defense gene annotations," "gene neighborhoods surrounding the most upregulated genes," or "regions shared among more than five species." This interactive functionality allows immediate visual inspection and iterative refinement of criteria, bridging the gap between data exploration and downstream analysis. Moreover, *gggenomes* enables the export of relevant genes and metadata for further analysis, supporting a more integrated and efficient comparative genomics workflow than other tools.

The utility of *gggenomes* has already been demonstrated across a range of studies. It was used to analyze mobile genetic elements in bacterial and eukaryotic genomes, enabling the efficient exploration of novel elements and their distribution across various environments in microbial ecology[2–4]. Furthermore, it facilitated the identification of expanding gene families in moth genomes and the examination of resistance loci in grapevine and fungal pathogens, creating new insights in the fields of genome evolution and biotechnology[5–7].

The package's adaptability also extends to large-scale studies of viromics and biosynthetic gene clusters. It was instrumental in visualizing novel bacteriophage clades as part of a large-scale expansion of the RNA virome[8], and it supports automated workflows for biosynthetic gene cluster analysis with BGCFlow[9] and exploring fungal giant mobile elements with STARFISH[10]. Collectively, these examples illustrate how *gggenomes* enhances research across a wide range of genomic research, from evolutionary and ecological genomics to virology and biotechnology.

In conclusion, *gggenomes* provides an intuitive, flexible, and effective solution for exploring and visualizing comparative genomic data. Building on the strengths of the R *tidyverse* and

*ggplot2*, it streamlines genomic workflows and helps researchers generate clear, informative visualizations that enhance the understanding of complex genomic relationships.

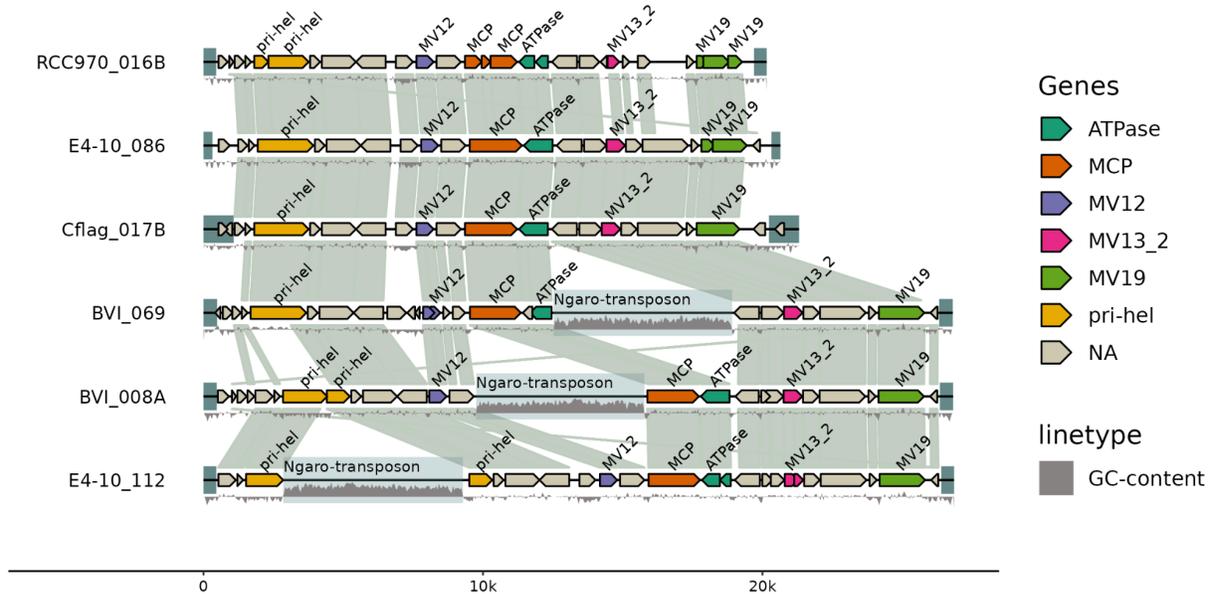

**Figure 1 | Comparative visualization of the genomic architectures of endogenous viruses and associated mobile genetic elements**. Shown is a gggenomes plot of six viral genomes[4], one per row, with genome names on the left. Annotated genes are depicted as arrows. Colors encode functionally conserved gene clusters, and functional labels are also shown above each gene. Terminal inverted repeats at the end of each genome are indicated by dark-blue boxes. Syntenic regions between neighboring genomes are indicated via gray blocks connecting areas of detectable nucleotide-level sequence similarity. The GC-content within sliding windows along each genome is indicated by ribbons below each genome sequence, normalized to the overall GC-content median. Ngaro transposons inserted into the viral genomes and overlapping with regions of high GC-content are labeled and marked by semitransparent dark-blue boxes.